\documentclass{WileyMSP-template}

\usepackage{chemformula} 
\usepackage[T1]{fontenc} 
\usepackage[separate-uncertainty=true, per-mode = symbol]{siunitx}
\usepackage{pdflscape}
\usepackage[compress]{cite}

\newcommand{\evs}{E_\mathrm{VS}}
\DeclareMathOperator{\erf}{erf}

\begin{document}

\title{Atomistic compositional details and their importance for spin qubits in isotope-purified silicon-germanium quantum wells }

\maketitle




\begin{affiliations}
Jan Klos\\
JARA-FIT Institute for Quantum Information, Forschungszentrum Jülich GmbH \& RWTH Aachen University, Aachen, Germany\\

J. Tröger\\
Institute of Materials Physics, University of Münster, Münster, Germany\\
and Tascon GmbH, Münster, Germany\\

Jens Keutgen\\
I. Physikalisches Institut IA, RWTH Aachen University, Aachen, Germany\\

Merritt P. Losert\\
University of Wisconsin-Madison, Madison, Wisconsin, USA\\

H. Riemann, Nikolay V. Abrosimov\\
Leibniz-Institut für Kristallzüchtung (IKZ), Berlin, Germany\\

Joachim Knoch\\
Institute of Semiconductor Electronics, RWTH Aachen University, Aachen, Germany\\

Hartmut Bracht\\
Institute of Materials Physics, University of Münster, Münster, Germany\\

Susan N. Coppersmith\\
University of New South Wales, Sydney, Australia\\

Mark Friesen\\
University of Wisconsin-Madison, Madison, Wisconsin, USA\\

Oana Cojocaru-Mirédin\\
I. Physikalisches Institut IA, RWTH Aachen University, Aachen, Germany and\\
INATECH, Albert-Ludwigs Universität Freiburg, Freiburg im Breisgau, Germany\\
\textit{Email Address: Oana.Cojocaru-Miredin@mail.inatech.uni-freiburg.de}\\

Lars R. Schreiber\\
JARA-FIT Institute for Quantum Information, Forschungszentrum Jülich GmbH \& RWTH Aachen University, Aachen, Germany and\\
ARQUE Systems GmbH, Aachen, Germany\\
\textit{Email Address: Schreibr@physik.rwth-aachen.de}\\

Dominique Bougeard\\
Institut für Experimentelle und Angewandte Physik, Universität Regensburg,
Regensburg, Germany\\
\textit{Email Address: Dominique.Bougeard@ur.de}\\
\end{affiliations}

\author{}


\newpage

\begin{abstract}

Understanding crystal characteristics down to the atomistic level increasingly emerges as a crucial insight for creating solid state platforms for qubits with reproducible and homogeneous properties. Here, isotope concentration depth profiles in a SiGe/$^{28}$Si/SiGe heterostructure are analyzed with atom probe tomography (APT) and time-of-flight secondary-ion mass spectrometry down to their respective limits of isotope concentrations and depth resolution. Spin-echo dephasing times $T_2^\mathbf{echo}=\SI{128}{\micro\second}$ and valley energy splittings $\evs$ around \SI{200}{\micro\electronvolt} have been observed for single spin qubits in this quantum well (QW) heterostructure, pointing towards the suppression of qubit decoherence through hyperfine interaction with crystal host nuclear spins or via scattering between valley states. The concentration of nuclear spin-carrying $^{29}$Si is \SI{50 \pm 20}{ppm} in the $^{28}$Si QW. The resolution limits of APT allow to uncover that both the SiGe/$^{28}$Si and the $^{28}$Si/SiGe interfaces of the QW are shaped by epitaxial growth front segregation signatures on a few monolayer scale. A subsequent thermal treatment, representative of the thermal budget experienced by the heterostructure during qubit device processing, broadens the top SiGe/$^{28}$Si QW interface by about two monolayers, while the width of the bottom $^{28}$Si/SiGe interface remains unchanged. Using a tight-binding model including SiGe alloy disorder, these experimental results suggest that the combination of the slightly thermally broadened top interface and of a minimal Ge concentration of $0.3 \%$ in the QW, resulting from segregation, is instrumental for the observed large $\evs=\SI{200}{\micro\electronvolt}$. Minimal Ge additions $< 1 \%$, which get more likely in thin QWs, will hence support high $\evs$ without compromising coherence times. At the same time, taking thermal treatments during device processing as well as the occurrence of crystal growth characteristics into account seems important for the design of reproducible qubit properties.

\end{abstract}

\section{Introduction}
The development of novel quantum technologies in condensed matter - and in particular the quest for a scalable and fault-tolerant quantum computer (QC) – increasingly ties decisive device performance parameters, such as the qubit fidelities and coherence times, to atomistic scale material properties. This strongly increases the need to get quantitative insights and, ideally, to find ways to control and custom-tailor atomistic details of condensed matter. Lately, obtaining chemical and spatial information at interfaces with atomistic precision has been pointed out to play a paramount role in the development of large-scale quantum information processors based on all major envisaged solid-state qubits such as superconducting qubits, spin qubits in quantum dots (QDs) or color centers, and topological qubits\cite{Frolov2020, Murray2021, Premkumar2021, Place2021}. 

 Thin-film epitaxially-grown $^{28}$Si/SiGe heterostructures consisting of a tensile strained, isotope-purified $^{28}$Si quantum well (QW) layer, sandwiched between two layers of the alloy SiGe, have been proven to be an excellent host for spin qubits \cite{Philips2022,Neyens2024}. These qubits are realized by controlling single to few electron spins in electrostatically-defined quantum dots in the QW \cite{Wild2012, Zwanenbourg2013}. Important ingredients for a fault-tolerant QC are long spin decoherence times, single and two-qubit gates as well as fast spin detection; all with fidelities beyond the quantum error correction threshold \cite{Yoneda2018,Struck2021, Kammerloher2021,Noiri2022,Mills2022, Xue2022}. The need of medium-distance quantum information transfer \cite{Vandersypen2017} triggered research on coherent transport of spin qubits using few operation signals \cite{Seidler2022, Xue2023, Struck2023}, which poses high demand on material homogeneity \cite{Langrock2023}. All these studies suggest that materials properties need to be understood on an atomistic scale to enable the realization of a $^{28}$Si/SiGe-based large-scale solid-state QC \cite{Kuenne2023}. In particular, the atomistic details of the semiconductor heterostructure seem to be highly relevant for two sources of qubit decoherence: hyperfine interaction of the free electron spin qubit with nuclear spins of the host crystal lattice and intervalley spin decoherence in the $^{28}$Si QW. The relevant hyperfine contact interaction depends on the concentration of lattice atoms carrying non-zero nuclear spins in the $^{28}$Si QW, with which the electron wavefunction of a considered spin qubit overlaps \cite{Assali2011}. This concentration is influenced by the degree of isotope purification in the $^{28}$Si QW, possibly by the $^{28}$Si/SiGe heterostructure epitaxy or also post-growth bulk diffusion processes, thermally triggered for example during the qubit device fabrication \cite{Neul2024}. The intervalley spin decoherence, on the other hand, is particularly relevant in the case of a non-desirable, uncontrolled occupation of an excited valley state during a spin qubit operation in the valley ground state \cite{Ferdous2018, Langrock2023, Losert2024}. The relevant metric is the valley-splitting energy ($\evs$) between the excited and the ground valley state, which needs to be sufficiently large for low-decoherence qubit operation. The atomistic details of $^{28}$Si/SiGe interfaces such as atomic steps and SiGe alloy disorder \cite{Ando1979, Boykin2004, Friesen2006, Friesen2007, Culcer2010, Yang2020, Hosseinkhani2020, Shi2011, Wuetz2021, Chen2021, Dodson2022, Losert2023, Pena2024} have been pointed out to be highly relevant for the magnitude of $\evs$ in a heterostructure, correlating with a significant spreading of experimentally determined valley energy splittings reported in the literature in various heterostructures \cite{Shi2011, Borselli2011, Kawakami2014, Scarlino2017, Zajac2015, Mi2017, Watson2018, Ferdous2018, Borjans2019, Hollmann2020, McJunkin2021, Mi2018, Wuetz2021, Esposti2024, Volmer2023}. Being able to analyze spatial depth concentration profiles down to the few atomistic monolayers and the few 100 ppm concentration level of isotopes in as-grown heterostructures as well as in processed devices is hence of importance to devise and test strategies to produce devices with well-controlled $\evs$. \\

Here, we analyze a Si$_{0.7}$Ge$_{0.3}$/$^{28}$Si/Si$_{0.7}$Ge$_{0.3}$ heterostructure grown by solid-source molecular beam epitaxy (MBE, see Experimental Section). Our focus lies on the depth-resolution of the composition profiles at the interfaces between $^{28}$Si and Si$_{0.7}$Ge$_{0.3}$ and on the isotope concentration of $^{28}$Si investigated with pulsed laser atom probe tomography (APT) and time-of-flight secondary-ion mass spectrometry (ToF-SIMS). We compare as-grown samples with post-growth annealed samples. The post-growth annealing is representative of the highest thermal budget used during qubit device processing (see Experimental Section). Spin qubit devices processed from the same Si$_{0.7}$Ge$_{0.3}$/$^{28}$Si/Si$_{0.7}$Ge$_{0.3}$ heterostructure used in the study presented here have previously been shown to feature excellent properties in terms of single spin qubit robustness, with a valley splitting energy $\evs$ ranging from \SI{185}{\micro\electronvolt} to \SI{212}{\micro\electronvolt} \cite{Hollmann2020}, an ensemble spin dephasing time $T_2^*\approx \SI{20}{\micro\second}$ and a spin-echo dephasing time $T_2^\mathbf{echo}=\SI{128}{\micro\second}$. Both dephasing times were not limited by the hyperfine contact interaction of residual $^{29}$Si isotopes in the QW \cite{Struck2020}.

We find < \SI{60}{ppm} nuclear spin-carrying $^{29}$Si in the $^{28}$Si QW by APT, confirming the absence of isotope diffusion during the epitaxy or post-growth anneal and in line with the qubit coherence times observed in devices made from this heterostructure. APT allows us to resolve a slight broadening of the top Si$_{0.7}$Ge$_{0.3}$/$^{28}$Si QW interface by approximately two monolayers (1 ML = 0.132\,nm) after a thermal anneal representative of device processing, compared to the as-grown interface. Furthermore, our analysis uncovers slight signatures of segregation that occurred at the crystal growth front during the epitaxy of both interfaces, the Si$_{0.7}$Ge$_{0.3}$/$^{28}$Si top and the $^{28}$Si/Si$_{0.7}$Ge$_{0.3}$ bottom interfaces. Moreover, our APT analysis suggests that the segregation at the bottom interface may result in the lowest Ge concentration reached in the \SI{10.5}{\nano\metre} thick $^{28}$Si QW to be at most $0.3\%$. Using a tight-binding and an effective-mass model, we find that slight details of the experimental Ge concentration profiles, such as the comparatively subtle effects of the post-growth annealing, seem to induce valley splitting energies around \SI{200}{\micro\electronvolt}.

\section{Concentration profile of the SiGe/$^{28}$Si/SiGe quantum well}

\subsection{Heterostructure characteristics and measurement conditions}

\begin{figure}
    \centering
    \includegraphics[width=8 cm]{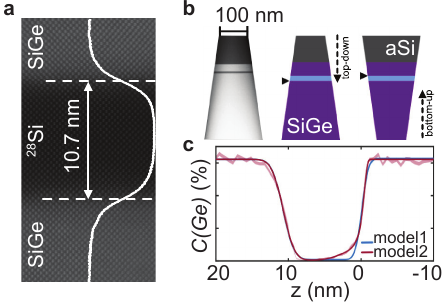}
    \caption{\textbf{a} High resolution HAADF STEM image of the $^{28}$Si/SiGe heterostructure with an intensity profile obtained on all detected atomic columns. 
    \textbf{b} APT tip and two schematic tip configurations used to probe the top interface (bottom-up) and bottom interface (top-down) of the QW.
    \textbf{c} Ge concentration profile obtained by APT across the QW, fitted by Eq. \ref{Eq1} (model1) and by Eq. \ref{Eq1} and Eq. \ref{Eq2} (model2) respectively (see Experimental Section).}
    \label{Fig:method}
\end{figure}

The Si$_{0.7}$Ge$_{0.3}$/$^{28}$Si/Si$_{0.7}$Ge$_{0.3}$ QW part of the heterostructure has been grown at a nominal temperature of \SI{350}{\celsius} with SiGe potential barriers of natural isotope composition and a QW highly purified in $^{28}$Si (see Experimental Section). Figure \ref{Fig:method}a shows a high-angle annular dark field scanning transmission electron micrograph (HAADF STEM) of the QW. The laterally averaged profile intensity (shown as an overlay in Figure \ref{Fig:method}a) and the performed ToF-SIMS analysis (see Supporting Information) agree in a measured QW thickness of \SI{10.5\pm0.2}{\nano\metre}. This value of the QW thickness was used in APT data analysis, in addition to the correction method by Vurpillot et al. \cite{Vurpillot2004} and the Landmark reconstruction \cite{Fletcher2020}, to obtain a precise 3D reconstruction of the heterostructure (see Experimental Section). For the Landmark reconstruction, a 15 atom-percent Ge isosurface is used.
Moreover, it has been shown \cite{Koelling2009, Dyck2017, Koelling2023} that collecting an APT profile from SiGe across an interface with Si is less precise than collecting it from Si across the interface with SiGe, because of the different evaporation fields of each material. Hence, to ensure highest resolution, we have produced dedicated specimen tips (Figure \ref{Fig:method}b) for the APT analysis of each QW interface, to always probe a heterostructure interface from Si across SiGe. Thus, the \textit{bottom interface} is analyzed using a specimen tip probed in a \textit{top-down}, while the \textit{top interface} is analyzed using a specimen tip probed in a \textit{bottom-up} configuration, as sketched by arrows in Figure \ref{Fig:method}b and conducted in Ref. \cite{Dyck2017}. We measured two tips of each type and compared the results with another set of, in total, four specimens, which we annealed prior to preparing the tips (see Experimental Section). The interface broadenings of the Si and Ge profiles at the top and bottom QW interfaces determined from each individual analyzed tip are provided in the Supporting Information.

To determine interface profile widths, we fit the experimental, depth-resolved concentration profiles with an error-function model which has been widely used for the analysis of isotope profiles \cite{Suedkamp2016, Kube2010, Kube2013}. For all TEM, ToF-SIMS and APT experimental profiles, we chose to fit the top and bottom interface together within one profile by using a model with two opposing error functions, to suppress numerical errors (see Experimental Section). As discussed below, for APT profiles, we extend the error function fit at the bottom interface to better take into account a self-limiting effect of Ge segregation during growth of a Si/SiGe interface \cite{Godbey1992,Fukatsu1991}. Both fit models are shown on an exemplary APT profile in Figure \ref{Fig:method}c and are discussed in the Experimental Section.

In the following, $C(X)$ denotes the depth-resolved composition profile of $X$, while $C(z; X)$ is the composition value of $X$ at the depth $z$ in growth direction. Along the manuscript, $X$ will either be \textit{Si} (standing here for the sum of the composition in the isotopes $^{29, 30}$Si), \textit{Ge} (standing for the sum of the composition in the isotopes $^{70, 72, 73, 74, 76}$Ge) or a specific isotope, in particular $^{28}Si$. Also, $r^\prime_\mathrm{t;b}(X)$ stands for the interface width of the profile of $X$ at the QW interface $y$ in the as-grown heterostructure, while $r_\mathrm{t;b}(X)$ represents the post-growth annealed counterpart, with $t$ for the \textit{top} or $b$ for the \textit{bottom} QW interface.

\subsection{Experimental analysis of the SiGe/$^{28}$Si top interface}

\begin{figure}
    \centering
    \includegraphics[width=12 cm]{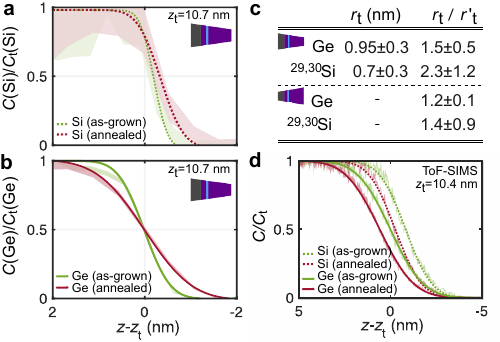}
    \caption{APT and ToF-SIMS measurements of the top SiGe/$^{28}$Si interface. Shaded areas in the profiles correspond to a single standard deviation interval of the measured data.  
    \textbf{a}: APT data and fit of the two isotopes $^{29,30}$Si concentration profile around the position of the top-interface $z_\mathrm{t}$ and normalized to the fitted Si concentration $C^\mathrm{Si}_t$ of the top-barrier. The inset sketches the bottom-up APT specimen tips used for a precise analysis of the top SiGe/$^{28}$Si interface (cf. Fig. \ref{Fig:method}c). \textbf{b}: APT data as in panel \textbf{a}, here for the concentration profile of Ge. 
    \textbf{c}: Table of the APT interface widths $r_\mathrm{t}$ for the annealed samples and of their ratio with the as-grown width $r^\prime_\mathrm{t}$. Upper half: Fit parameters resulting from data obtained in bottom-up analysis, which is more precise for this top QW interface. Lower half: Comparison to ratio values obtained in top-down analysis, which is less precise for this top QW interface. 
    \textbf{d}: Corresponding concentration profiles measured by ToF-SIMS.}
    \label{Fig:topinterface}
\end{figure}

Figure \ref{Fig:topinterface} displays the depth-resolved composition profiles across the SiGe/$^{28}$Si top interface by APT and ToF-SIMS and the corresponding parts of the fits to the profiles. For highest resolution of this top interface, we plot results of the APT specimens analyzed in bottom-up configuration.

We first consider the depth-resolved concentration profile $C(\mathrm{Si})$ along the growth direction: Figure \ref{Fig:topinterface}a shows this profile for the as-grown and for the post-growth annealed samples respectively. The profile of the as-grown sample is well fitted by our error-function model (see Eq. \ref{Eq1} in the Experimental Section). The fit yields an interface width $r^\prime_\mathrm{t}(\mathrm{Si})=\SI{0.31 \pm 0.09}{\nano\metre}$ which corresponds to $2.4$ monolayers of $^{28}$Si tensile strained by relaxed Si$_{0.7}$Ge$_{0.3}$. We postulate that $r^\prime_\mathrm{t}(\mathrm{Si})=\SI{0.31 \pm 0.09}{\nano\metre}$ is approaching the resolution limit of our APT measurement and of the 3D reconstruction, because an efficient mechanism which may cause a significant broadening of the top Si composition interface profile is not evident: The presence of segregation at the growth front during epitaxy will only concern subsurface Ge atoms, which may exchange with surface Si atoms, but there is no driving force for up-floating of Si atoms in the growth direction \cite{Fukatsu1991}. Also, there is no driving force for spontaneous intermixing of Si isotopes across a $^{28}$Si/$^{nat}$SiGe interface. Furthermore no significant thermally-driven diffusion is to be expected at the growth temperature of \SI{350}{\celsius} \cite{Suedkamp2016, Kube2010}.

Figure \ref{Fig:topinterface}b depicts the Ge profile $C(\mathrm{Ge})$ around the top QW interface in the as-grown and in the annealed samples. With $r^\prime_\mathrm{t}(\mathrm{Ge})=\SI{0.65 \pm 0.07}{\nano\metre}$ it is slightly larger than the interface width $r^\prime_\mathrm{t}(Si)$ discussed in Fig. \ref{Fig:topinterface}a. We attribute this interface broadening to the segregation of Ge atoms, expected for a SiGe growth front overgrowing Si (here $^{28}$Si) in ultra high vacuum epitaxy, termed as \textit{leading edge} in the literature \cite{Godbey1992,Fukatsu1991, Godbey1994, Godbey1998, Jernigan1996, Jernigan1997}. Considering $r^\prime_\mathrm{t}(\mathrm{Si})=\SI{0.31 \pm 0.09}{\nano\metre}$ to approach the resolution limit of our APT profiles, as discussed above, this broadening of the interface due to Ge segregation may be as low as $r_\mathrm{t}^{\prime}(\mathrm{Ge})-r_\mathrm{t}^{\prime}(\mathrm{Si})=\SI{0.34 \pm 0.16}{\nano\metre}$, which corresponds to $2.3$ monolayers of relaxed Si$_{0.7}$Ge$_{0.3}$. The interface width of the APT Ge profile is comparable to recent APT studies of as-grown structures produced in chemical vapor deposition (CVD) \cite{Dyck2017,Wuetz2021, Koelling2023}. Note that segregation may be hampered under certain CVD growth conditions and that segregation is not discussed in these latter works.

In addition to the slight broadening of the Ge profile, we observe a second evidence of Ge leading edge segregation at the as-grown SiGe/$^{28}$Si top interface of the QW: the turning point of the Ge profile ($C_\mathrm{Ge}/C^{\mathrm{Ge}}_t=0.5$) is slightly shifted compared to the Si profile, as a consequence of the tendency of Ge atoms to float upon the growth front, up to a certain equilibrium concentration \cite{Fukatsu1991}. Comparing Figure \ref{Fig:topinterface}a and Figure \ref{Fig:topinterface}b, the average shift is \SI{0.16 \pm 0.11}{\nano\metre} for the APT profiles of as-grown samples. To our knowledge, the experimental observation of this shifted turning point between Ge and Si in the presence of leading edge segregation has not been reported before.

For comparison to the as-grown Si and Ge profiles, Figs. \ref{Fig:topinterface}a and \ref{Fig:topinterface}b also show the profiles of post-growth annealed samples, which represent a realistic thermal budget during qubit device processing (see Experimental Section). The Ge and Si profiles clearly  reveal a broadening at the interface which we attribute to an isotropic bulk diffusion of Ge and Si in the heterostructure crystal during post-growth annealing. The average interface width of the Ge profile is $r_\mathrm{t}(\mathrm{Ge})=\SI{0.95 \pm 0.3}{\nano\metre}$, showing a larger variation between the results of the two specimens ($r_\mathrm{t}(\mathrm{Ge})=[\SI{0.76 \pm 0.03}{\nano\metre}, \SI{1.13 \pm 0.05}{\nano\metre}]$, see Supporting Information) than for the as-grown specimen. The interfacial broadening of the Si profile is $r_\mathrm{t}(\mathrm{Si})=\SI{0.7 \pm 0.3}{\nano\metre}$, again with some variation among the two specimens ($r_\mathrm{t}=[\SI{0.5 \pm 0.1}{\nano\metre}, \SI{0.8 \pm 0.1}{\nano\metre}]$, see Supporting Information). If we consider $r^\prime_\mathrm{t}(\mathrm{Si})=\SI{0.31 \pm 0.09}{\nano\metre}$ to approach the resolution limit of our APT analysis, as discussed earlier, $r_\mathrm{t}(\mathrm{Ge})-r_\mathrm{t}^{\prime}(\mathrm{Si})=\SI{0.7 \pm 0.5}{\nano\metre}$ represents the broadening of the Ge profile due to post-growth annealing. The table in Figure \ref{Fig:topinterface}c summarizes the interface widths $r_\mathrm{t}$ after the post-growth anneal and the ratio $r_\mathrm{t}/r^\prime_\mathrm{t}$ between post-growth annealed and as-grown samples. As it is expected for the bulk diffusion in the heterostructure crystal, we find similar values for the ratios $r_\mathrm{t}/r^\prime_\mathrm{t}$ for Ge and Si. In the lower half of the table in in Figure \ref{Fig:topinterface}c, we also mention the ratios $r_\mathrm{t}/r^\prime_\mathrm{t}$ determined on APT needles analyzed in the top-down configuration. Although these conditions are less precise for the analysis of the top QW interface, the ratios for Ge and Si confirm the broadening of the top QW interfaces.

Figure \ref{Fig:topinterface}d shows the Ge and Si profiles acquired by ToF-SIMS, both for as-grown and post-growth annealed samples. The fitted interface widths $r_\mathrm{t}$ and $r^\prime_\mathrm{t}$ for Ge and Si with values higher than \SI{1.5}{\nano\metre} significantly exceed the interface widths obtained from APT. No significant broadening of the interface by annealing is evident in ToF-SIMS, demonstrating a lower depth resolution compared to APT. However, ToF-SIMS offers a better signal-to-noise ratio for individual data points of the concentration profile than APT due to the larger probe area of $\SI{100}{\micro\metre} \times \SI{100}{\micro\metre}$, which we integrate to calculate $C(z, Ge)$ and $C(z, Si)$. Notably, the probe area of APT corresponds to the size of a quantum dot\cite{Hollmann2020,Struck2020} and is thus representative for the environment to which a single spin qubit is exposed. In the ToF-SIMS crater, a RMS roughness of \SI{3.2}{\nano\metre} is detected by atom force microscopy (see Supporting Information). This roughness dominates the interface widths in the measured concentration profiles. We attribute a significant part of this roughness in the crater to the presence of cross-hatching in this type of heterostructure as discussed in the Supporting Information. The cross-hatching, which manifests as a regular terracing on a \si{\micro\metre} scale in two perpendicular crystal directions, results from the strain relaxation via dislocations in the concentration-graded buffer part of the heterostructure and is, hence, unavoidable. It thus seems that we have reached the resolution limit of ToF-SIMS in terms of the quantification of the interface width in this type of heterostructure, presumably due to cross-hatching in our layer structure. Note that also the HAADF-STEM analysis results in a broader interface width compared to APT (Fig. \ref{Fig:method}).

Summarizing the comparison between as-grown and post-growth annealed samples, we observe a post-growth diffusional broadening of the Ge and Si profiles caused by annealing. The slight profile broadening is resolved by means of APT but not by ToF-SIMS measurements. Both APT and ToF-SIMS reveal signatures of segregation of Ge during the growth process, which becomes evident in a slightly retarded onset of the Ge compared to the Si profile for the top Si$_{0.7}$Ge$_{0.3}$/$^{28}$Si interface.

\subsection{Experimental analysis of the $^{28}$Si/SiGe bottom interface}

\begin{figure}
\centering
    \includegraphics[width=12 cm]{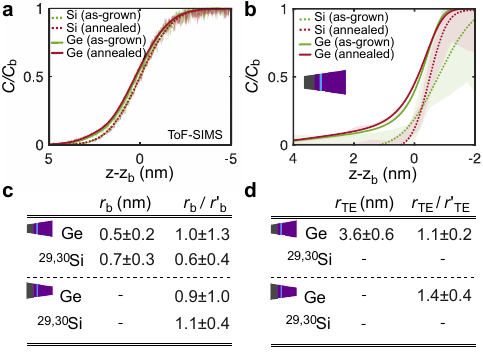}
    \caption{APT and ToF-SIMS analysis of the bottom interface. Shaded areas in the profiles correspond to a single standard deviation interval of the measured data. 
    \textbf{a: }ToF-SIMS data for Si and Ge of the bottom QW interface for as-grown and annealed samples.  
    \textbf{b:} APT data for Si and Ge of the bottom QW interface for as-grown and annealed samples. The inset sketches the analysis conditions in the top-down configuration, more precise for the bottom QW interface. 
    \textbf{c:} Table of the APT interface widths $r_\mathrm{b}$ for the annealed samples and of their ratio with the as-grown width $r^\prime_\mathrm{b}$. Upper half: Fit parameters resulting from data obtained in top-down analysis, which is more precise for this bottom QW interface. Lower half: Comparison to ratio values obtained in bottom-up analysis, which is less precise for this bottom QW interface.
    \textbf{d:} Same as in \textbf{c} for the Ge trailing edge width $r_\mathrm{TE}$ (see Eq. \ref{Eq2}).
   }
    \label{Fig:bottominterface}
\end{figure}

Figure \ref{Fig:bottominterface} displays the depth-resolved composition profiles across the $^{28}$Si/Si$_{0.7}$Ge$_{0.3}$ bottom QW interface by ToF-SIMS and APT as well as the corresponding fits to the profiles. For highest resolution of this interface, we we plot results of the APT specimens analyzed in top-down configuration.

Figure \ref{Fig:bottominterface}a shows the Ge- and the Si profiles acquired by ToF-SIMS, both for as-grown and post-growth annealed samples. We find these profiles to be very similar to those observed for the top interface in Figure \ref{Fig:topinterface}d. Indeed, all profiles are accurately described with a simple error-function, following the model given by Eq. \ref{Eq1} in the Experimental Section. Note, however, that for a Si overgrowth of Ge or SiGe, a segregation of Ge atoms into the Si overgrowth layer has been reported and experimentally resolved for as-grown MBE structures grown at higher substrate temperatures than our heterostructure in the past \cite{Godbey1992,Fukatsu1991, Godbey1998}. During Si overgrowth of SiGe, a self-limiting mechanism of Ge segregation at the growth front is invoked \cite{Fukatsu1991}, leading to a stretched Ge concentration profile, termed as \textit{trailing edge} in the literature \cite{Godbey1992}. We conclude that our experiments show that a Ge segregation and in particular its self-limiting character for $^{28}$Si overgrowth of SiGe at the bottom interface cannot be resolved via ToF-SIMS for this QW grown at \SI{350}{\celsius}.

In contrast, a stretched Ge profile at the QW bottom interface is clearly resolved by means of APT as shown in Figure \ref{Fig:bottominterface}b. While the Si profiles $C(\mathrm{Si})$ are accurately described with the error-function of Eq. \ref{Eq1} (see Experimental Section), this is not the case for the Ge profiles $C(\mathrm{Ge})$. After a drop in Ge concentration, from a certain threshold value, the decrease in the Ge concentration is less pronounced. This is indicative of a self-limitation of Ge segregation with increasing Si neighbouring and corresponds to the trailing edge discussed in the literature. To our knowledge, this is the most narrow trailing edge (as a consequence of the comparatively low substrate temperature in the epitaxy) that has been experimentally resolved. The comparison between ToF-SIMS and APT shows that such narrow trailing edges are only resolved in APT in these strain-relaxed heterostructures. To accurately describe the trailing edge, we extend our error function model, as discussed in Eq. \ref{Eq2} in the Experimental Section. In addition to the interface width $r^\prime_\mathrm{b}(\mathrm{Ge})$, we introduce a characteristic length $r^\prime_\mathrm{TE}(\mathrm{Ge})$, to quantify the self-limited segregation in the region of the trailing edge. For the as-grown structures our Ge trailing edge profiles are in line with model predictions for segregation in ultra high vacuum epitaxy at substrate temperatures of \SI{350}{\celsius} \cite{Fukatsu1991}. Note that from the fit of the experimental APT Ge concentration profile as parameterized by Eq. \ref{Eq1} and Eq. \ref{Eq2}, we find a non-zero minimum Ge concentration $C_{min}(\mathrm{Ge})$ in the QW: From $C_0(\mathrm{Ge})$ we deduce $C_{min}(\mathrm{Ge})=0.21 \%$ in the as-grown heterostructure, which we interpret as an additional manifestation, here in our fit model, of the presence of the segregation trailing edge.

To analyze the effect of post-growth annealing, we summarize the values of $r_\mathrm{b}$ and the ratios $r_\mathrm{b}/r^\prime_\mathrm{b}$ for Ge and Si in the table of Figure \ref{Fig:bottominterface}c as well as $r_\mathrm{TE}$ and $r_\mathrm{TE}/r^\prime_\mathrm{TE}$ for Ge in the table of Figure \ref{Fig:bottominterface}d. We also include the ratios determined from the less precise \cite{Dyck2017} APT specimen configuration (here bottom-up tips) to increase the statistics in the determination of the ratios. In contrast to the clear trend observed in the APT profiles of the Ge and Si at the top interfaces, the interfacial broadening of Si and Ge at the bottom interface is not significantly affected by post-growth  annealing: The standard deviations of the as-grown and of the annealed data overlap in the whole range of analysis.  Regarding the additional fit parameter $C_0(\mathrm{Ge})$ (Eq. \ref{Eq1}), we deduce a value of $C_{min}(\mathrm{Ge})=0.42 \%$ in the annealed heterostructure, compared to $C_{min}(\mathrm{Ge})=0.21 \%$ in the as-grown structure, indicative of very slight bulk diffusion during the post growth anneal. Note that the ToF-SIMS analysis also does not display any difference between the as-grown and the post-growth annealed samples, as is expected, given the roughness-limited depth resolution addressed in the analysis of the top interface.

\subsection{Quantification of the Si isotope and of the minimal Ge concentration in the QW}
Going beyond the capabilities of APT in terms of depth resolution, we also explore the limits of the method in terms of the minimal resolvable Si isotope composition and minimal Ge composition in the $^{28}$Si QW. To ensure a high accuracy in the determination of the isotope composition, we increased the sensitivity to detect $^{29}$Si and $^{30}$Si ions inside the $^{28}$Si QW by employing a horizontal APT needle configuration (see Experimental Section) instead of the bottom-up or top-down geometries used before. We also describe the calibration of our measurement in the Experimental Section, using a piece of the $^{28}$Si MBE crystal source as a reference and the $^{14.5}$Si/$^{14}$Si ratio of the double charged isotopes to avoid an interference with the mass-to-charge state $29$ induced by $^{28}$SiH.

We find a composition of \SI{50 \pm 20}{ppm} of $^{29}$Si in the QW by APT. This is in excellent agreement with the \SI{41}{ppm} composition we determined for the $^{28}$Si MBE source crystal and suggests that no modification or contamination of the Si isotope enrichment takes place during the evaporation of the source crystal in MBE \cite{Suedkamp2016, Kube2010}. The composition of the $^{30}$Si isotope drops to \SI{10}{ppm} inside the $^{28}$Si layer. In the source crystal it was previously determined to be \SI{1.3}{ppm}(=1.3$\times $10$^{-4}$ {\%}), in a different measurement \cite{Pramann2011}. We conclude that \SI{10}{ppm} represents our detection limit of APT for Si isotopes in the QW layer. This proves the high detection capability of APT even in  very confined space such as a \SI{10.5}{\nano\metre} thick QW layer.

The non-annealed APT needle in horizontal configuration is also advantageous to determine the minimal detectable Ge concentration in the QW. We determine a Ge concentration of \SI{0.3}{\%}, which we consider as the upper boundary of the Ge concentration within the \SI{10.5}{\nano\metre} QW, considering the detection limit of APT for the horizontal needle configuration. This experimental value is in good agreement with the fit results for the minimum Ge concentration in the QW $C_{min}(\mathrm{Ge})$ deduced from $C_0(\mathrm{Ge})$ in Eq. \ref{Eq1} for the as-grown ($C_{min}(\mathrm{Ge})=0.21 \%$) and the annealed ($C_{min}(\mathrm{Ge})=0.42 \%$) samples.

\section{Valley splitting estimates for the realistic Ge concentration profiles}
The heterostructure profiles determined by APT and ToF-SIMS describe Ge concentrations $C(\mathrm{Ge})$ as a function of depth along the growth direction $z$, spatially averaging in the analyzed plane $(x;y)$ at a given depth $z$. The analyzed area $(x;y)$ depends on the method, APT or ToF-SIMS. Note that the random nature of the SiGe alloy causes these profiles to vary spatially in plane, on an atomistic level, which in turn may cause fluctuations of the valley splitting. The mean and variance of the valley splitting both increase sensitively when the electron wavefunction overlaps strongly with Ge~\cite{Losert2023}, as we will illustrate below. Given this sensitivity of the valley splitting and the fact that the typical confinement area of current experimentally implemented spin qubits approach the analysis area of APT, a series of statistically meaningful measurements of the valley splitting on a piece of a given heterostructure therefore, in principle, could provide a sensitive probe of the Ge concentration, even in the low-Ge regime where the detection limit of APT measurements is reached. Up to now, such statistically meaningful samplings of the valley splitting have not been realized experimentally.
Concerning the heterostructure studied here, a number of measurements were previously obtained in one device \cite{Hollmann2020, Struck2020} from the same annealed heterostructure studied here, yielding values in the range of $\evs=185$-$\SI{212}{\micro\electronvolt}$, as the center position of the dot was shifted by 6~nm and, additionally, $\evs>$\SI{100}{\micro\electronvolt} estimated for the few other devices produced from that heterostructure. 
Below, we show that (i) these relatively high valley-splitting values and their variations are consistent with theoretical predictions for the annealed sample, and (ii) a non-vanishing value of $C_{min}^{Ge}$ is crucial for obtaining the observed results.
We also perform large-scale valley-splitting simulations to demonstrate how subtle features in the Ge concentration can have strong effects on the valley splitting.

To begin, we adopt the minimal 1D two-band tight-binding model of Boykin et al. \cite{Boykin2004}, which has been shown to quantitatively predict $\evs$ behavior in real devices \cite{Wuetz2021,Losert2023}. To model alloy disorder in this 1D geometry, we start with the APT Ge concentration profiles of as-grown and annealed samples fitted with Eqs. \ref{Eq1} and \ref{Eq2}. We then introduce small random fluctuations in the Ge concentration for each atomic layer, consistent with random alloy disorder, following the approach of Ref.~\cite{Losert2023}. To build up a large statistical sample, we repeat this randomization many times and simulate $\evs$ for each case. We also compare the distributions of tight-binding simulation results to effective-mass theory, which predicts Rayleigh distributions for disorder-dominated valley splittings \cite{Wuetz2021, Losert2023}. More details on the theoretical tools used here are presented in the Supporting Information.

\begin{figure}
\centering
    \includegraphics[width= 17 cm]{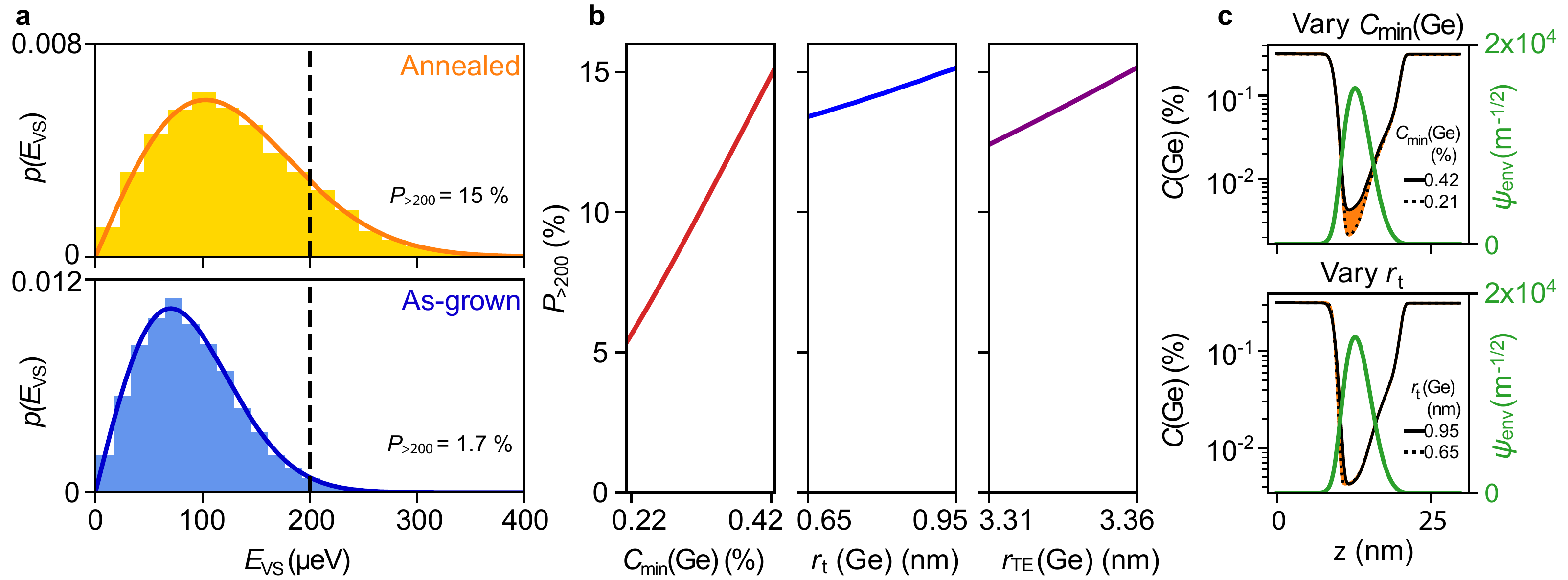}
    \caption{Valley splitting simulations of the as-grown and post growth annealed quantum wells illustrate the impact of a small amount of Ge in the quantum well. \textbf{a} Histograms show 10,000 simulations of $\evs$ for the annealed (top) and as-grown (bottom) quantum-well profiles, obtained using a minimal two-band tight-binding model \cite{Boykin2004}. Each simulation represents a different realization of alloy disorder. Solid lines indicate the expected Rayleigh distribution for $\evs$, derived from effective-mass theory \cite{Losert2023}. $P_{>200}$ indicates the probability of obtaining $\evs$ larger than~\SI{200}{\micro\electronvolt} (dashed line), as derived from these distributions. \textbf{b} Starting with the annealed quantum well profile, we vary the minimum Ge concentration in the well, $C_{min}(\mathrm{Ge})$, from its value in the annealed well ($C_{min}(\mathrm{Ge}) \approx 0.42$), to its value in the as-grown well, ($C_{min}(\mathrm{Ge}) \approx 0.21$), and we compute the resulting $P_{>200}$ values (left, red). As the Ge content is reduced, so is $P_{>200}$. We perform the same analysis with the top interface width $r_t(\mathrm{Ge})$ (center, blue) and the trailing edge width $r_{TE}(\mathrm{Ge})$ (right, purple). \textbf{c} The annealed quantum well profile (black solid lines) is compared with two modified quantum wells (black dotted lines). Top: starting with the annealed well, the modified quantum well is found by reducing $C_{min}(\mathrm{Ge})$ from 0.42 to 0.21. Bottom: we reduce the top interface width $r_t(\mathrm{Ge})$ from 0.95 to \SI{0.65}{\nano\meter}. In both plots, we highlight the Ge concentration differential in orange. We also include a simulation of the 1D quantum dot wavefunction envelope, $\psi_{env}$ (green).}
    \label{Fig:Fig4}
\end{figure}

In Fig.~\ref{Fig:Fig4}a, we show histograms of $10,000$ 1D tight-binding simulations of $\evs$ for the annealed QW (top panel) and the as-grown QW (bottom panel). The solid lines show the corresponding Rayleigh distributions derived from the effective-mass theory, using the respective QW parameters as inputs. 
We first confirm that the experimentally measured valley splittings, with typical values of $\SI{200}{\micro\electronvolt}$, are realistic and expected for this system. To do this, we compute the probability of finding valley splittings larger than this value, $P_{>200} = P(\evs >$ \SI{200}{\micro\electronvolt}), for both of these distributions, where \SI{200}{\micro\electronvolt} is indicated in the figure by the dashed line. 
This analysis suggests that \SI{200}{\micro\electronvolt} is on the high side of the predicted distribution, but not unreasonably so. Moreover, we see that $P_{>200}$ is more than eight times larger for the annealed sample than for the as-grown sample, highlighting the strong dependence of the valley splitting on details of the Ge concentration profile and emphasizing the potential impact of sample processing on the valley splitting values determined in spin qubit experiments.

To illustrate this further, in Fig.~\ref{Fig:Fig4}b we study the effect of three fitting parameters in Eq.~\ref{Eq1} and Eq.~\ref{Eq2} on the valley splitting distribution: the minimum Ge concentration in the QW $C_{min}(\mathrm{Ge})$, the width of the top interface $r_\text{t}(\mathrm{Ge})$, and the trailing edge parameter $r_\text{TE}(\mathrm{Ge})$. 
In each case, to ensure physically reasonable results, we ramp a single parameter between its as-grown and its post-anneal values, while keeping the other parameters fixed at their values for the annealed samples. 
The effect of such variations on the Ge density profile is subtle, as illustrated in Fig.~\ref{Fig:Fig4}c for the $C_{min}(\mathrm{Ge})$ and $r_t(\mathrm{Ge})$ parameters (note the logarithmic scale).
Indeed, such variations approach the resolution limit of our fitting procedure (see Supporting Information).
However, these variations significantly affect the valley splitting. Even slight, experimentally barely resolvable variations of $C_{min}(\mathrm{Ge})$ between $0.22$ and $0.42 \%$ have a particularly strong effect, since they increase the Ge concentration in the region where the dot wavefunction is large, as shown in Fig.~\ref{Fig:Fig4}c.
Fig. \ref{Fig:Fig4}b and \ref{Fig:Fig4}b illustrate that variations of $r_\text{T}$ and $r_{TE}$ yield smaller, but non-negligible contributions. This analysis confirms the previous claims that, (i) for heterostructures without super-sharp features (defined as features sharper than 2-3 atomic monolayers), fluctuations arising from alloy disorder dominate the valley splitting, and (ii) this effect is enhanced when the wavefunction overlaps more strongly with the Ge~\cite{Losert2023}. 
In the Supporting Information, we also show that experimentally observed variations of $\evs$ between 185 and $\SI{212}{\micro\electronvolt}$ \cite{Hollmann2020} are consistent with dot-center shifts of $\SI{6}{\nano\metre}$, for the $C_{min}(\mathrm{Ge})$ levels considered here.

\section{Implications for spin qubits}
Our tight-binding and effective mass models, using the experimental APT Ge profiles as input, give strong indications that $\evs$ is dominated by alloy disorder in our heterostructure, as in others recently studied with APT \cite{Dyck2017,Wuetz2021} or scanning probe methods \cite{Pena2024}. Clearly, post-growth annealing during the device fabrication, possibly enhanced by the presence of a slight segregation trailing edge at the bottom of the QW, seems to mostly be responsible for the experimental observation of valley splitting energies up to \SI{212}{\micro\electronvolt} \cite{Hollmann2020} and $>$\SI{100}{\micro\electronvolt} for the few other spin qubit devices experimentally tested in this heterostructure. Our study indicates that the ability to experimentally determine monolayer-scale details in the Ge concentration QW profile of fabricated devices to capture post-growth annealing effects, such as with APT, will be instrumental in providing input parameters for quantitative modelling of $\evs$, as well as for developing epitaxy recipes to maximize $\evs$. The knowledge of the top and bottom QW interface widths is not sufficient to capture the relevant effects. Any QW profile fit needs to be complemented by parameters reflecting the Ge environment of the wavefunction down to concentrations below $0.5 \%$, as illustrated by the parameter $C_{min}(\mathrm{Ge})=0.42 \%$ in our study. Particularly noteworthy, the thinner the $^{28}$Si QW, the higher the probability that the wavefunction will overlap a Ge concentration relevant enough to boost $\evs$ in heterostructures with alloy disorder-dominated interfaces, as shown in Fig.~\ref{Fig:Fig4}c. Interestingly, recent reports on spin qubit devices fabricated in heterostructures with comparatively thin QWs frequently report experimentally determined $\evs>$ \SI{100}{\micro\electronvolt} \cite{Chen2021, Wuetz2021, Esposti2024}.

Our model predicts $P_{>100} = 62 \%$ for the annealed samples. Experimentally, we and the other recent reports on comparatively thin CVD grown QWs \cite{Chen2021, Wuetz2021, Esposti2024} have found $\evs>\SI{100}{\micro\electronvolt}$ in each measured quantum dot device. There are also reports where, in each measured qubit position, the valley splitting is smaller than $\evs<\SI{100}{\micro\electronvolt}$ \cite{Volmer2023}. Although the amount of measured devices from each heterostructure is not sufficient to be statistically meaningful, these observations from different groups may hint that additional parameters are relevant for quantitative modelling, such as spatial correlations in the Ge concentration fluctuations and local strain in the QW \cite{Corley2023a, Corley2023b}. Dislocation networks may produce such features (and also induce cross-hatching of the surface). The ability to conduct experimental $\evs$ mappings in more extended quantum dot devices, such as electron shuttlers \cite{Volmer2023}, should soon allow to further experimentally test the predicted variability of $\evs$ in heterostructures with alloy disorder-dominated interfaces.
 
Regarding the dephasing times of the spin qubit, we obtained high and charge noise-limited dephasing times $T_2^*\approx \SI{20}{\micro\second}$ \cite{Struck2020}, a record spin-echo dephasing time $T_2^\mathbf{echo}=\SI{128}{\micro\second}$ \cite{Struck2020} and an electron g-factor of $g=2.00 \pm 0.01$ \cite{Hollmann2020} for single spin qubits in our heterostructure. This indicates that neither the segregation-induced, slightly delayed onset of Ge compared to natural Si at the top interface, nor the weak signatures of post-growth annealing-induced diffusion within the whole QW impact the spin qubit coherence in this $^{28}$Si QW. Our results thus suggest that small additions of Ge (below $0.5 \%$ in our heterostructure) may provide a balance between a sufficiently large valley-splitting $\evs$ for spin qubit manipulation without introducing uncontrolled dephasing due to hyperfine interaction with $^{73}$Ge nuclear spins, which is much larger per atom than the one with $^{29}$Si~\cite{Wilson1964} nuclear spins. Anticipating further improvements in coherence times, isotope-purified Ge could then be used in and around the $^{28}$Si QW \cite{Sailer2009}, suppressing hyperfine interaction with $^{73}$Ge. Also, the fact that we determined $g=2.00 \pm 0.01$ \cite{Hollmann2020} in this heterostructure, demonstrates that the slight addition of Ge to the $^{28}$Si QW does not induce relevant spin-orbit coupling.

\section{Conclusion}
Summarizing our findings, our analysis allows us to experimentally extract realistic concentration profiles of the MBE heterostructure in its as-grown state as well as after a post-growth thermal anneal, representative of the thermal budget to which a spin qubit device is exposed during sample processing. We have found state of the art width values for the as-grown interfaces between the $^{28}$Si QW and the top and bottom Si$_{0.7}$Ge$_{0.3}$ barrier. The post-growth anneal leads to a seemingly small, but clearly detectable, broadening of the top interface due to isotropic bulk diffusion, while the bottom barrier width remains unchanged within the experimental detection limit. At the same time, we reveal signs of Ge segregation on comparably small length scales, imputable to the growth of the heterostructure. Segregation trailing edges have been reported on significantly larger length scales (due to the use of higher substrate temperatures during the epitaxy) before \cite{Fukatsu1991}. APT proves here to be highly suited to analyze such rather subtle signatures of segregation and post-growth annealing. In comparison, we found ToF-SIMS to reveal the slightly retarded turn-on of Ge at the top interface, but to be resolution-limited regarding the trailing edge at the bottom interface and also the effect of post-growth annealing. We attribute the resolution limitation to the heterostructure-inherent strain relaxation.
Using an APT needle in horizontal configuration to increase the analysis volume of the \SI{10.5}{\nano\metre} thin QW, APT allowed us to assess an isotope purity of the $^{28}$Si QW of \SI{50\pm20}{ppm} $^{29}$Si and down to the detection limit of \SI{10}{ppm} for $^{30}$Si. Additionally, we found the upper boundary of the Ge composition within the $^{28}$Si QW to be \SI{0.3}{\%} in the horizontal needle configuration, which represents an unprecedented level of precision in a segregation or diffusion study in Si. This upper boundary agrees well with the parameter $C_{min}(\mathrm{Ge})$ of our APT Ge concentration profile fit function which increases from $0.21 \%$ in the as-grown to $0.42 \%$ in the annealed QW and suggests that post-growth annealing may have slightly increased the minimum Ge concentration present in the QW, presumably as a consequence of annealing-induced post-growth bulk diffusion.

Our theoretical model, which uses the fits to the experimental APT Ge concentration profiles as an input, provides strong indications that it is actually the post-growth annealing, representative of the maximum thermal budget applied during quantum dot device processing, that is responsible for the experimental observation of large valley splitting energies for the spin qubits tested in this heterostructure. Indeed, the experimentally observed and comparably small increase of two fit function parameters under post-growth annealing - the width of the top barrier and a concentration offset suggesting a non-zero minimum Ge concentration around $0.3 \%$ in the QW - suffices to boost the probability to find $\evs>$\SI{200}{\micro\electronvolt} by more than a factor of 8 in the model. Our study strongly points out the importance of subtle Ge concentration changes in the direct environment of each qubit wavefunction. Notably, our results suggest that the risk of finding particularly low $\evs$ may be significantly reduced in qubit devices fabricated in comparatively thin QW heterostructures in the regime of alloy fluctuation-dominated QW interfaces, in correlation with recent experimental studies \cite{Chen2021, Wuetz2021, Esposti2024}.    

Hence, by employing the outstanding resolution limits of APT in Si/SiGe - in terms of depth resolution and also composition resolution in a nanometer-scale probe volume - we show that being able to experimentally determine realistic concentration profiles down to the few-monolayer-limit and to concentrations $< 1 \%$ allows to resolve signatures of thin film growth-inherent phenomena like Ge segregation or slight post-growth anneal-induced diffusion on such low length scales. Given that thin QWs, atomically sharp interfaces, delta-like Ge spikes, sharp superlattices or the addition of a precise concentration of Ge to the QW are envisaged \cite{Losert2023} as an ingredient for massive scalability of spin qubits in Si/SiGe, empirical knowledge on such phenomena at realistic sample fabrication conditions will be key to develop viable novel heterostructure design approaches.

\section{Experimental Section}

\threesubsection{Molecular beam epitaxy}\\
All heterostructures are grown in a solid source molecular beam epitaxy (MBE) ultra high vacuum chamber equipped with three independent electron beam evaporators for Si and Ge of natural isotopic composition and isotope-purified $^{28}$Si with a 41\,ppm residual composition of $^{29}Si$ determined in APT.
At first, a relaxed Si$_{1-x}$Ge$_x$ virtual substrate is grown on a Si(100) substrate without intentional miscut with natural isotopic composition and increasing Ge composition $x$ up to a target composition of $x=30 \%$. The growth temperature for the virtual substrate is \SI{500}{\celsius}.
Next, \SI{300}{\nano\metre} Si$_{0.7}$Ge$_{0.3}$ is grown followed by a nominal \SI{12}{\nano\metre} strained QW layer of $^{28}$Si. The growth temperature for the QW is \SI{350}{\celsius} with a deposition rate of \SI{0.14}{\angstrom\per\second}.
Finally, a \SI{45}{\nano\metre} relaxed Si$_{0.7}$Ge$_{0.3}$ layer is grown, capped by nominally \SI{1.5}{\nano\metre} of natural Si.

Post-growth thermal treatment is done using a rapid thermal process (RTP) for \SI{15}{\second} at \SI{700}{\celsius} with a \SI{5}{\kelvin\per\second} ramp on a SiC carrier, referred to as the annealed heterostructure.

\threesubsection{Atom probe tomography}\\
For atom probe tomography (APT), an additional \SI{200}{\nano\metre} of electron beam evaporated amorphous silicon is deposited onto the SiGe heterostructure to prevent excessive damage during focused ion beam (FIB) processing. All APT measurements were performed on a Cameca local-electrode atom-probe (LEAP) 4000X-Si system with a picosecond ultraviolet (wavelength of \SI{355}{\nano\metre}) laser. The experimental conditions are given by a base temperature of \SI{30}{\kelvin}, pulse repetition rate of \SI{250}{\kilo\hertz}, detection rate of \SI{1}{\%} and laser energy of \SI{30}{\pico\ampere}. The sample reconstruction was done using the software IVAS. The STEM analysis was performed on a TITAN (S)TEM from FEI using an accelerating voltage of \SI{200}{\kilo\volt}. The sample was rotated in edge on condition and viewed along the <110> direction. The HAADF-STEM images were processed using Gatan’s digital micrograph software.

\begin{figure}
\centering
    \includegraphics[width=11cm]{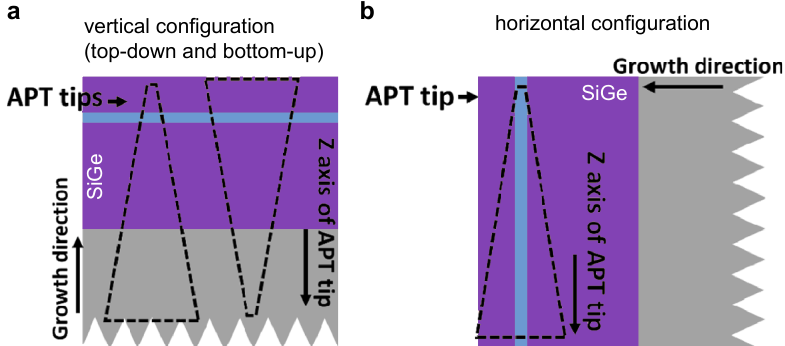}
    \caption{Schematic of the two tip geometries used for the APT analysis of the Si$_{0.7}$Ge$_{0.3}$/$^{28}$Si/Si$_{0.7}$Ge$_{0.3}$ heterostructure: \textbf{a: } Vertical geometry. It delivers the tips for our two different analysis configurations for the isotope depth profiles in the QW, as highlighted by the dashed line: top-down (left) and bottom-up (right), see also Fig. \ref{Fig:method}b. \textbf{b:}  Horizontal geometry. 
   }
    \label{Fig:FigS1}
\end{figure}

Prior to APT measurements, the needle-shape specimens are prepared using the dual-beam FIB system (FEI Helios NanoLab 650i) and annular Ga$^{+}$ milling. 
As has been reported, interface aberrations occur during APT measurements for materials with changing evaporation fields \cite{Dyck2017}. To infer these deviating effects, we prepare two specimen tips with opposing direction as schematically shown in Figure Fig. \ref{Fig:FigS1}a and Fig. \ref{Fig:method}b to individually probe the top (bottom-up specimen tip) and bottom (top-down specimen tip) QW interface. We refer to both as the vertical needle configuration (Fig. \ref{Fig:FigS1}a). All specimen tips yield an approximate diameter of \SI{100}{\nano\metre}. The 3D specimen reconstruction is calibrated with high resolution scanning transmission electron microscopy (HR STEM) measurements of the QW thickness in the prepared specimen tips.

\begin{figure}
\centering
    \includegraphics[width=8.5cm]{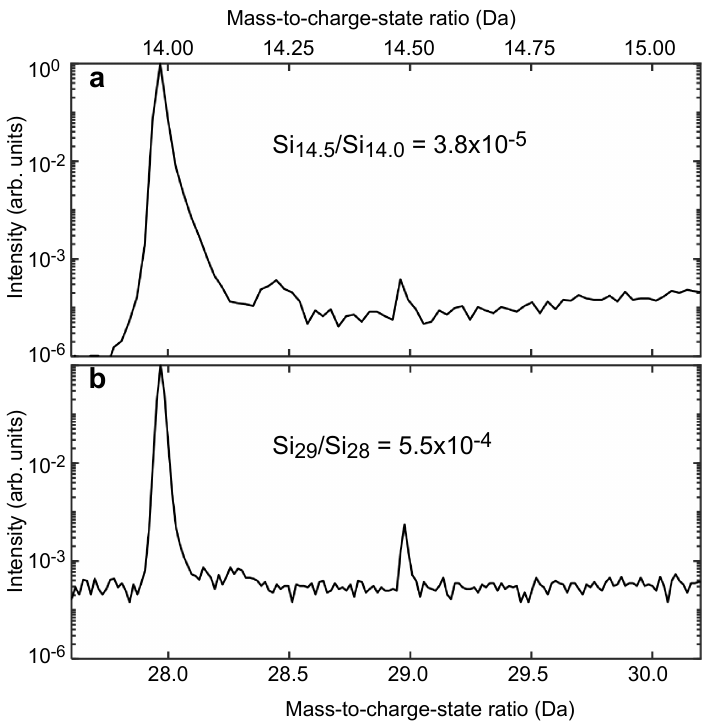}
    \caption{APT mass spectrum of the 99.9957 pure $^{28}$Si single crystal reference sample. \textbf{a} Double charged Si ions peaks Si$_{14.5}$ and Si$_{14}$ . \textbf{b} Single charged Si ion peaks Si$_{29}$ and Si$_{28}$ mass peaks.
   }
    \label{Fig:FigS3}
\end{figure}

We increased the sensitivity of APT to spurious concentrations of Ge and Si isotopes in the QW by preparing a needle in a horizontal configuration (Fig. \ref{Fig:FigS1}b). This allows to integrate a larger volume of atoms detected in the middle of the QW. \cite{Cojocaru2013}
We measure a \SI{50\pm20}{ppm} of residual $^{29}$Si in the geometric center of the QW using mass-to-charge conversion based on single and double charged Si isotopes as explained below. In order to quantify the $^{29}$Si composition within the $^{28}$Si QW, we have used a crystalline piece of the $^{28}$Si MBE source material (99.9957\,\% $^{28}$Si single crystal with 41\,ppm of $^{29}$Si \cite{Becker2010}) as a reference sample. The mass spectra obtained for both, single and double charged Si isotopes, are shown in Fig.~\ref{Fig:FigS3}. Both spectra reveal a discrepancy between the ratios of Si$_{14.5}$ over Si$_{14}$ mass peaks (value is $3.8 \times 10^{-5}$) and ratio of Si$_{29}$ over Si$_{28}$ mass peaks (value is $5.5\times 10^{-4}$). If we calculate the nominal $^{29}$Si over $^{28}$Si isotope concentration ratio in the 99.9957 \% pure $^{28}$Si single crystal, we obtain a value of $4.1\times 10^{-5}$. This is almost the same as the value calculated for the Si$_{14.5}$ over Si$_{14}$ mass peak ratio in Fig. \ref{Fig:FigS3}a. Thus, this APT determined ratio is correct, while the Si$_{29}$ peak is overestimated by APT due to the overlap with the SiH peak. Therefore, the correct  $^{29}$Si composition determined by APT will be given by:
$$
 ^{29}\mathrm{Si}=\left(1+ \frac{\mathrm{Si}_{28}}{\mathrm{Si}_{14}} \right) \mathrm{Si}_{14.5}  
$$
Applied to the APT data performed on the 99.9957 pure $^{28}$Si single crystal reference sample, this expression yields a value of $41\pm10$\,ppm which fits exactly to previous measurements \cite{Becker2010}.

\begin{figure}
\centering
    \includegraphics[width= 8 cm]{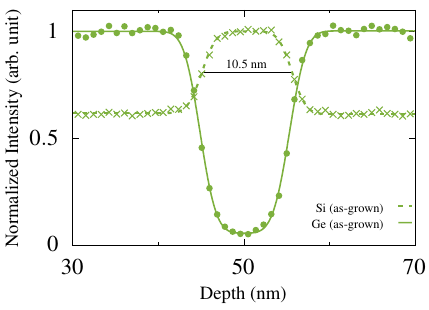}
    \caption{Normalized ToF-SIMS intensities of $^{28}$Si and $^{70,72,73,74,76}$Ge (sum of all Ge isotopes) as a function of depth in the QW region (every 20th data point shown). By fitting Eq. \ref{Eq1} to the profiles, the width of the QW was determined to be \SI{10.5\pm0.1}{\nano\metre}.}
    \label{Fig:SIMS_QW-Thickness}
\end{figure}

\threesubsection{Time-of-Flight Secondary Ion Mass Spectrometry ToF-SIMS}\\
The SIMS experiments were performed using an ION-ToF ToF.SIMS 5 system operating in dual beam mode. While a first ion beam (\SI{500}{\electronvolt} O$_2^{+}$, \SI{40}{\nano\ampere}) was sputtering a crater with a base area of $\SI{300}{\micro\metre} \times \SI{300}{\micro\metre}$, a second ion beam (\SI{15}{\kilo\electronvolt} Bi$_1^{+}$, \SI{0.25}{\pico\ampere}) was progressively analysing an area of $\SI{75}{\micro\metre} \times \SI{75}{\micro\metre}$ in the center of the crater bottom. For optimum oxidation of the sample, oxygen flooding with a partial pressure of \SI{1E-6}{\milli\bar} was used. The depths of the sputtered craters were determined with an uncertainty of about \SI{10}{\%} using a Bruker DektakXT mechanical profilometer.\newline
Fig. \ref{Fig:SIMS_QW-Thickness} shows the normalized intensities of $^{28}$Si and $^{70, 72, 73, 74, 76}$Ge (sum of all Ge isotopes) vs. depth in the QW region. By fitting Eq. \ref{Eq1} the width of the QW was determined to be \SI{10.5\pm0.1}{\nano\metre}.

\threesubsection{Fit model for the Si and Ge concentration at the interfaces}\\
For the analysis of the measured isotopic profiles, models based on error functions and sigmoid functions have been widely used \cite{Dyck2017, Wuetz2021, Suedkamp2016, Kube2010, Kube2013}. We use an error function based concentration profile $C(z)$ in this work, to describe both heterostructure QW interfaces (the \textit{top interface} and the \textit{bottom interface}) for $\mathrm{X}=\mathrm{Si}$ or Ge:
\begin{equation}\label{Eq1}
    C(z;X)=\dfrac{C_t}{2} \cdot \bigg ( \erf{\left(-\dfrac{z-z_\mathrm{t}}{r_\mathrm{t}(X)}\right)} + 1\bigg )
    + \dfrac{C_b}{2}\cdot \bigg (\erf{\left(\dfrac{z-z_\mathrm{b}}{r_\mathrm{b}(X)}\right)}+1\bigg )
    +C_0(X)
\end{equation}
\noindent where $z_\mathrm{t}$ and $z_\mathrm{b}$ are the positions of the top and bottom interface along the growth z-axis (defined by the concentration profile of Ge), respectively. The corresponding interface width can be different in general and is given by $r_\mathrm{t}$ and $r_\mathrm{b}$, respectively. The concentration steps at the interfaces are given by $C_t$ and $C_b$ for the top and bottom interface, respectively. We also take a constant offset concentrations of $C_0$ into account. Note that $C_0 \leq C_\mathrm{min}$, where $C_\mathrm{min}$ is the minimum concentration of the profile, depending on $C_0$ but also on the broadening of the interfaces and the quantum well width. This model assumes a single constant parameter to describe the interface width. Depending upon the growth conditions, more complex profiles such as trailing edges can emerge, when Ge segregation affects the Ge incorporation rate at the heterostructure growth front. Therefore, we extend the model to capture such effects, which we dominantly observed for the bottom interface.  We suggest to use a non-constant bottom interface width $r_\mathrm{b}(z; Ge)$ in Eq.\ref{Eq1}: 
\begin{equation}\label{Eq2}
    r_\mathrm{b}(z; Ge)=\dfrac{r_\mathrm{TE}(\mathrm{Ge})-r_\mathrm{b}(\mathrm{Ge})}{2}\cdot \left( \erf{\left(-\dfrac{z-z_\mathrm{b}-\epsilon}{r_\epsilon}\right)}+1 \right) +r_\mathrm{b}(\mathrm{Ge}).
\end{equation}
Here, we assume two regimes of different Ge incorporation rates split by a specific Ge concentration threshold $C(z_\mathrm{b}+\epsilon; Ge)$ on the wafer surface during growth, where $\epsilon \ll r_\mathrm{b}$. For $C(z;Ge)\gg C(z_\mathrm{b}+\epsilon; Ge)$, the interface width is described by $r_\mathrm{b}$ equivalent to the model in Equation \ref{Eq1}. For $C(z;Ge)\ll C(z_\mathrm{b}+\epsilon; Ge)$, the interface width is $r_\mathrm{TE}>r_\mathrm{b}$ due to the decreased Ge incorporation rate. For changing Ge concentrations approaching $C(z_\mathrm{b}\pm \epsilon; Ge)$, we assume a Gaussian transition from $r_\mathrm{b}(\mathrm{Ge})$ to $r_\mathrm{TE}(\mathrm{Ge})$ or vice versa within $0\leq r_\epsilon \leq 2.5$ nm. The latter limit is used to ensure smooth transitions between both regimes and corresponds to the width of the assumed Gaussian across the overall wafer.\newline

Finally, we provide a comparison of the interface width parameters $r_\mathrm{t}$ and $r_\mathrm{b}$ of our error function model with the sigmoid model used in the literature for recent analyses of CVD grown heterostructures \cite{Dyck2017, Wuetz2021}:
\begin{equation}
    \dfrac{2}{1+e^{-\left(z-z_0\right)/\tau}}-1 \approx \dfrac{x-z_0}{2\tau}+\mathcal{O}\left(\left(\dfrac{x-z_0}{2\tau}\right)^3\right)
\end{equation}
where $z$ is the position along the growth direction, $z_0$ is the position of the material interface and $\tau$ is the interface diffusion length in the sigmoid model \cite{Dyck2017, Wuetz2021}.
The used error function model for a single material interface is described by
\begin{equation}
    \erf{\left(\dfrac{x-z_0}{r}\right)}\approx 
    \dfrac{2}{\sqrt{\pi}}\dfrac{x-z_0}{r}+\mathcal{O}\left(\left(\dfrac{x-z_0}{r}\right)^3\right)
\end{equation}
Hence, the relation between the interface widths respectively defined in both models is $4\tau \approx\sqrt{\pi}r$. We use this relation for making comparisons in the table summarizing all measured APT needles in the Supporting Information.

\medskip
\textbf{Supporting Information} \par
Supporting Information is available 
from the authors.

\medskip
\textbf{Acknowledgements} \par 
This work was funded in part by the German Research Foundation (DFG) within the projects 421769186 (SCHR 1404/5-1), 289786932 (BO 3140/4-2 and SCHR 1404/3-2) and Germany's Excellence Strategy - Cluster of Excellence Matter and Light for Quantum Computing" (ML4Q) EXC 2004/1 - 390534769 as well as by the Federal Ministry of Education and Research under Contract No. FKZ: 13N14778. Project Si-QuBus received funding from the QuantERA ERA-NET Cofund in Quantum Technologies implemented within the European Union's Horizon 2020 Programme. The research was also sponsored by the Army Research Office (ARO) under Awards No.\ W911NF-22-1-0090 and W911NF-23-1-0110.
The views, conclusions, and recommendations contained in this document are those of the authors and are not necessarily endorsed nor should they be interpreted as representing the official policies, either expressed or implied, of the Army Research Office (ARO) or the U.S. Government. The U.S. Government is authorized to reproduce and distribute reprints for Government purposes notwithstanding any copyright notation herein.

\medskip

%
\bibliographystyle{MSP}
\bibliography{references}

\begin{thebibliography}{10}
\providecommand{\url}[1]{\texttt{#1}}
\providecommand{\urlprefix}{URL }

\bibitem{Frolov2020}
S.~M. Frolov, M.~J. Manfra, J.~D. Sau,
\newblock \emph{Nat. Phys.} \textbf{2020}, \emph{16} 718.

\bibitem{Murray2021}
C.~E. Murray,
\newblock \emph{Mater. Sci. Eng. R Rep} \textbf{2021}, \emph{146} 100646.

\bibitem{Premkumar2021}
A.~Premkumar, C.~Weiland, S.~Hwang, B.~J{\"a}ck, A.~P.~M. Place, I.~Waluyo,
  A.~Hunt, V.~Bisogni, J.~Pelliciari, A.~Barbour, M.~S. Miller, P.~Russo,
  F.~Camino, K.~Kisslinger, X.~Tong, M.~S. Hybertsen, A.~A. Houck, I.~Jarrige,
\newblock \emph{Commun. Mater.} \textbf{2021}, \emph{2} 72.

\bibitem{Place2021}
A.~P.~M. Place, L.~V. Rodgers, P.~Mundada, B.~M. Smitham, M.~Fitzpatrick,
  Z.~Leng, A.~Premkumar, J.~Bryon, A.~Vrajitoarea, S.~Sussman, G.~Cheng,
  T.~Madhavan, H.~K. Babla, X.~H. Le, Y.~Gang, B.~J{\"a}ck, A.~Gyenis, N.~Yao,
  R.~J. Cava, N.~P. {de Leon}, A.~A. Houck,
\newblock \emph{Nat. Commun.} \textbf{2021}, \emph{12} 1779.

\bibitem{Philips2022}
S.~G.~J. Philips, M.~T. M{\k{a}}dzik, S.~V. Amitonov, S.~L. de~Snoo, M.~Russ,
  N.~Kalhor, C.~Volk, W.~I.~L. Lawrie, D.~Brousse, L.~Tryputen, B.~P. Wuetz,
  A.~Sammak, M.~Veldhorst, G.~Scappucci, L.~M.~K. Vandersypen,
\newblock \emph{Nature} \textbf{2022}, \emph{609} 919.

\bibitem{Neyens2024}
S.~Neyens, O.~K. Zietz, T.~F. Watson, F.~Luthi, A.~Nethwewala, H.~C. George,
  E.~Henry, M.~Islam, A.~J. Wagner, F.~Borjans, E.~J. Connors, J.~Corrigan,
  M.~J. Curry, D.~Keith, R.~Kotlyar, L.~F. Lampert, M.~T. Mądzik, K.~Millard,
  F.~A. Mohiyaddin, S.~Pellerano, R.~Pillarisetty, M.~Ramsey, R.~Savytskyy,
  S.~Schaal, G.~Zheng, J.~Ziegler, N.~C. Bishop, S.~Bojarski, J.~Roberts, J.~S.
  Clarke,
\newblock \emph{Nature} \textbf{2024}, \emph{629} 80.

\bibitem{Wild2012}
A.~Wild, J.~Kierig, J.~Sailer, J.~W. Ager, E.~Haller, G.~Abstreiter, S.~Ludwig,
  D.~Bougeard,
\newblock \emph{Appl. Phys. Lett.} \textbf{2012}, \emph{100} 143110.

\bibitem{Zwanenbourg2013}
F.~A. Zwanenburg, A.~S. Dzurak, A.~Morello, M.~Y. Simmons, L.~C.~L. Hollenberg,
  G.~Klimeck, S.~Rogge, S.~N. Coppersmith, M.~A. Eriksson,
\newblock \emph{Rev. of Mod. Phys.} \textbf{2013}, \emph{85} 961.

\bibitem{Yoneda2018}
J.~Yoneda, K.~Takeda, T.~Otsuka, T.~Nakajima, M.~R. Delbecq, G.~Allison,
  T.~Honda, T.~Kodera, S.~Oda, Y.~Hoshi, N.~Usami, K.~M. Itoh, S.~Tarucha,
\newblock \emph{Nat. Nanotechnol.} \textbf{2018}, \emph{13} 102.

\bibitem{Struck2021}
T.~Struck, J.~Lindner, A.~Hollmann, F.~Schauer, A.~Schmidbauer, D.~Bougeard,
  L.~R. Schreiber,
\newblock \emph{Sci. Rep.} \textbf{2021}, \emph{11} 16203.

\bibitem{Kammerloher2021}
E.~Kammerloher, A.~Schmidbauer, L.~Diebel, I.~Seidler, M.~Neul, M.~Künne,
  A.~Ludwig, J.~Ritzmann, A.~Wieck, D.~Bougeard, L.~R. Schreiber, H.~Bluhm,
\newblock Preprint available at https://arxiv.org/abs/2107.13598,
  \textbf{2021}.

\bibitem{Noiri2022}
A.~Noiri, K.~Takeda, T.~Nakajima, T.~Kobayashi, A.~Sammak, G.~Scappucci,
  S.~Tarucha,
\newblock \emph{Nature} \textbf{2022}, \emph{601} 338.

\bibitem{Mills2022}
A.~R. Mills, C.~R. Guinn, M.~J. Gullans, A.~J. Sigillito, M.~M. Feldman,
  E.~Nielsen, J.~R. Petta,
\newblock \emph{Sci. Adv.} \textbf{2022}, \emph{8} eabn5130.

\bibitem{Xue2022}
X.~Xue, M.~Russ, N.~Samkharadze, B.~Undseth, A.~Sammak, G.~Scappucci, L.~M.~K.
  Vandersypen,
\newblock \emph{Nature} \textbf{2022}, \emph{601} 343.

\bibitem{Vandersypen2017}
L.~M.~K. Vandersypen, H.~Bluhm, J.~Clarke, A.~Dzurak, R.~Ishihara, A.~Morello,
  D.~Reilly, L.~Schreiber, M.~Veldhorst,
\newblock \emph{npj Quantum Inf.} \textbf{2017}, \emph{3} 34.

\bibitem{Seidler2022}
I.~Seidler, T.~Struck, R.~Xue, N.~Focke, S.~Trellenkamp, H.~Bluhm, L.~R.
  Schreiber,
\newblock \emph{npj Quantum Inf.} \textbf{2022}, \emph{8} 100.

\bibitem{Xue2023}
R.~Xue, M.~Beer, I.~Seidler, S.~Humpohl, J.~Tu, T.~Stefan, T.~Struck, H.~Bluhm,
  L.~R. Schreiber,
\newblock \emph{Nat. Commun.} \textbf{2024}, \emph{15} 2296.

\bibitem{Struck2023}
T.~Struck, M.~Volmer, L.~Visser, T.~Offermann, R.~Xue, J.-S. Tu,
  S.~Trellenkamp, {\L}.~Cywi\ifmmode~\acute{n}\else \'{n}\fi{}ski, H.~Bluhm,
  L.~R. Schreiber,
\newblock \emph{Nat. Commun.} \textbf{2024}, \emph{15} 1325.

\bibitem{Langrock2023}
V.~Langrock, J.~A. Krzywda, N.~Focke, I.~Seidler, L.~R. Schreiber,
  {\L}.~Cywi{\'n}ski,
\newblock \emph{PRX Quantum} \textbf{2023}, \emph{4} 020305.

\bibitem{Kuenne2023}
M.~K\"unne, A.~Willmes, M.~Oberl\"ander, C.~Gorjaew, J.~D. Teske, H.~Bhardwaj,
  M.~Beer, E.~Kammerloher, R.~Otten, I.~Seidler, R.~Xue, L.~R. Schreiber,
  H.~Bluhm,
\newblock Preprint available at https://arxiv.org/abs/2306.16348,
  \textbf{2023}.

\bibitem{Assali2011}
L.~V.~C. Assali, H.~M. Petrilli, R.~B. Capaz, B.~Koiller, X.~Hu, S.~Das~Sarma,
\newblock \emph{Phys. Rev. B} \textbf{2011}, \emph{83} 165301.

\bibitem{Neul2024}
M.~Neul, I.~V. Sprave, L.~K. Diebel, L.~G. Zinkl, F.~Fuchs, Y.~Yamamoto,
  C.~Vedder, D.~Bougeard, L.~R. Schreiber,
\newblock \emph{Phys. Rev. Mater.} \textbf{2024}, \emph{8} 043801.

\bibitem{Ferdous2018}
R.~Ferdous, E.~Kawakami, P.~Scarlino, M.~P. Nowak, D.~Ward, D.~Savage,
  M.~Lagally, S.~Coppersmith, M.~Friesen, M.~A. Eriksson, L.~M.~K. Vandersypen,
\newblock \emph{npj Quantum Inf.} \textbf{2018}, \emph{4} 26.

\bibitem{Losert2024}
M.~P. Losert, M.~Oberländer, J.~D. Teske, M.~Volmer, L.~R. Schreiber,
  H.~Bluhm, S.~N. Coppersmith, M.~Friesen,
\newblock Preprint available at https://arxiv.org/abs/2405.01832 (2024),
  \textbf{2024}.

\bibitem{Ando1979}
T.~Ando,
\newblock \emph{Physical Review B} \textbf{1979}, \emph{19} 3089.

\bibitem{Boykin2004}
T.~B. Boykin, G.~Klimeck, M.~A. Eriksson, M.~Friesen, S.~N. Coppersmith, P.~von
  Allmen, F.~Oyafuso, S.~Lee,
\newblock \emph{Appl. Phys. Lett.} \textbf{2004}, \emph{84} 115.

\bibitem{Friesen2006}
M.~Friesen, M.~A. Eriksson, S.~N. Coppersmith,
\newblock \emph{Appl. Phys. Lett.} \textbf{2006}, \emph{89} 202106.

\bibitem{Friesen2007}
M.~Friesen, S.~Chutia, C.~Tahan, S.~N. Coppersmith,
\newblock \emph{Phys. Rev. B} \textbf{2007}, \emph{75} 115318.

\bibitem{Culcer2010}
D.~Culcer, X.~Hu, S.~Das~Sarma,
\newblock \emph{Phys. Rev. B} \textbf{2010}, \emph{82} 205315.

\bibitem{Yang2020}
C.~H. Yang, R.~C.~C. Leon, J.~C.~C. Hwang, A.~Saraiva, T.~Tanttu, W.~Huang,
  J.~{Camirand Lemyre}, K.~W. Chan, L.~Y. Tan, F.~E. Hudson, K.~M. Itoh,
  A.~Morello, M.~Pioro-Ladri{\`{e}}re, A.~Laucht, A.~S. Dzurak,
\newblock \emph{Nature} \textbf{2020}, \emph{580} 350.

\bibitem{Hosseinkhani2020}
A.~Hosseinkhani, G.~Burkard,
\newblock \emph{Phys. Rev. Res.} \textbf{2020}, \emph{2} 043180.

\bibitem{Shi2011}
Z.~Shi, C.~B. Simmons, J.~R. Prance, J.~King~Gamble, M.~Friesen, D.~E. Savage,
  M.~G. Lagally, S.~N. Coppersmith, M.~A. Eriksson,
\newblock \emph{Appl. Phys. Lett.} \textbf{2011}, \emph{99} 233108.

\bibitem{Wuetz2021}
B.~P. Wuetz, M.~P. Losert, S.~Koelling, L.~E.~A. Stehouwer, A.-M.~J. Zwerver,
  S.~G.~J. Philips, M.~T. Mądzik, X.~Xue, G.~Zheng, M.~Lodari, S.~V. Amitonov,
  N.~Samkharadze, A.~Sammak, L.~M.~K. Vandersypen, R.~Rahman, S.~N.
  Coppersmith, O.~Moutanabbir, M.~Friesen, G.~Scappucci,
\newblock \emph{Nat. Commun.} \textbf{2022}, \emph{13} 7730.

\bibitem{Chen2021}
E.~H. Chen, K.~Raach, A.~Pan, A.~A. Kiselev, E.~Acuna, J.~Z. Blumoff,
  T.~Brecht, M.~D. Choi, W.~Ha, D.~R. Hulbert, M.~P. Jura, T.~E. Keating,
  R.~Noah, B.~Sun, B.~J. Thomas, M.~G. Borselli, C.~Jackson, M.~T. Rakher,
  R.~S. Ross,
\newblock \emph{Phys. Rev. Appl.} \textbf{2021}, \emph{15} 044033.

\bibitem{Dodson2022}
J.~P. Dodson, H.~E. Ercan, J.~Corrigan, M.~P. Losert, N.~Holman, T.~McJunkin,
  L.~F. Edge, M.~Friesen, S.~N. Coppersmith, M.~A. Eriksson,
\newblock \emph{Phys. Rev. Lett.} \textbf{2022}, \emph{128} 146802.

\bibitem{Losert2023}
M.~P. Losert, M.~A. Eriksson, R.~Joynt, R.~Rahman, G.~Scappucci, S.~N.
  Coppersmith, M.~Friesen,
\newblock \emph{Phys. Rev. B} \textbf{2023}, \emph{108} 125405.

\bibitem{Pena2024}
L.~F. Peña, J.~C. Koepke, J.~H. Dycus, A.~Mounce, A.~D. Baczewski, N.~T.
  Jacobson, E.~Bussmann,
\newblock \emph{npj Quantum Inf.} \textbf{2024}, \emph{10} 33.

\bibitem{Borselli2011}
M.~G. Borselli, R.~S. Ross, A.~A. Kiselev, E.~T. Croke, K.~S. Holabird, P.~W.
  Deelman, L.~D. Warren, I.~Alvarado-Rodriguez, I.~Milosavljevic, F.~C. Ku,
  W.~S. Wong, A.~E. Schmitz, M.~Sokolich, M.~F. Gyure, A.~T. Hunter,
\newblock \emph{Appl. Phys. Lett.} \textbf{2011}, \emph{98} 123118.

\bibitem{Kawakami2014}
E.~Kawakami, P.~Scarlino, D.~R. Ward, F.~R. Braakman, D.~E. Savage, M.~G.
  Lagally, M.~Friesen, S.~N. Coppersmith, M.~A. Eriksson, L.~M.~K. Vandersypen,
\newblock \emph{Nat. Nanotechnol.} \textbf{2014}, \emph{9} 666.

\bibitem{Scarlino2017}
P.~Scarlino, E.~Kawakami, T.~Jullien, D.~R. Ward, D.~E. Savage, M.~G. Lagally,
  M.~Friesen, S.~N. Coppersmith, M.~A. Eriksson, L.~M.~K. Vandersypen,
\newblock \emph{Phys. Rev. B} \textbf{2017}, \emph{95} 165429.

\bibitem{Zajac2015}
D.~M. Zajac, T.~M. Hazard, X.~Mi, K.~Wang, J.~R. Petta,
\newblock \emph{Appl. Phys. Lett.} \textbf{2015}, \emph{106} 223507.

\bibitem{Mi2017}
X.~Mi, C.~G. P\'eterfalvi, G.~Burkard, J.~R. Petta,
\newblock \emph{Phys. Rev. Lett.} \textbf{2017}, \emph{119} 176803.

\bibitem{Watson2018}
T.~F. Watson, S.~G.~J. Philips, E.~Kawakami, D.~R. Ward, P.~Scarlino,
  M.~Veldhorst, D.~E. Savage, M.~G. Lagally, M.~Friesen, S.~N. Coppersmith,
  M.~A. Eriksson, L.~M.~K. Vandersypen,
\newblock \emph{Nature} \textbf{2018}, \emph{555} 633.

\bibitem{Borjans2019}
F.~Borjans, D.~M. Zajac, T.~M. Hazard, J.~R. Petta,
\newblock \emph{Phys. Rev. Appl.} \textbf{2019}, \emph{11} 044063.

\bibitem{Hollmann2020}
A.~Hollmann, T.~Struck, V.~Langrock, A.~Schmidbauer, F.~Schauer, T.~Leonhardt,
  K.~Sawano, H.~Riemann, N.~V. Abrosimov, D.~Bougeard, L.~R. Schreiber,
\newblock \emph{Phys. Rev. Appl.} \textbf{2020}, \emph{13} 034068.

\bibitem{McJunkin2021}
T.~McJunkin, E.~R. MacQuarrie, L.~Tom, S.~F. Neyens, J.~P. Dodson,
  B.~Thorgrimsson, J.~Corrigan, H.~E. Ercan, D.~E. Savage, M.~G. Lagally,
  R.~Joynt, S.~N. Coppersmith, M.~Friesen, M.~A. Eriksson,
\newblock \emph{Phys. Rev. B} \textbf{2021}, \emph{104} 085406.

\bibitem{Mi2018}
X.~Mi, M.~Benito, S.~Putz, D.~M. Zajac, J.~M. Taylor, G.~Burkard, J.~R. Petta,
\newblock \emph{Nature} \textbf{2018}, \emph{555} 599.

\bibitem{Esposti2024}
D.~Degli~Esposti, L.~E.~A. Stehouwer, O.~Gül, N.~Samkharadze, C.~Déprez,
  M.~Meyer, I.~N. Meijer, L.~Tryputen, S.~Karwal, M.~Botifoll, J.~Arbiol, S.~V.
  Amitonov, L.~M.~K. Vandersypen, A.~Sammak, M.~Veldhorst, G.~Scappucci,
\newblock \emph{npj Quantum Inf.} \textbf{2024}, \emph{10} 32.

\bibitem{Volmer2023}
M.~Volmer, T.~Struck, A.~Sala, B.~Chen, M.~Oberländer, T.~Offermann, R.~Xue,
  L.~Visser, J.-S. Tu, S.~Trellenkamp, Łukasz Cywiński, H.~Bluhm, L.~R.
  Schreiber,
\newblock Preprint available at https://arxiv.org/abs/2312.17694,
  \textbf{2023}.

\bibitem{Struck2020}
T.~Struck, A.~Hollmann, F.~Schauer, O.~Fedorets, A.~Schmidbauer, K.~Sawano,
  H.~Riemann, N.~V. Abrosimov, {\L}.~Cywiński, D.~Bougeard, L.~R. Schreiber,
\newblock \emph{npj Quantum Inf.} \textbf{2020}, \emph{6} 2056.

\bibitem{Vurpillot2004}
F.~Vurpillot, D.~Larson, A.~Cerezo,
\newblock \emph{Surf. Interface Anal.} \textbf{2004}, \emph{36} 552.

\bibitem{Fletcher2020}
C.~Fletcher, M.~P. Moody, D.~Haley,
\newblock \emph{J. Phys. D: Appl. Phys.} \textbf{2020}, \emph{53} 475303.

\bibitem{Koelling2009}
S.~Koelling, M.~Gilbert, J.~Goossens, A.~Hikavyy, O.~Richard, W.~Vandervorst,
\newblock \emph{Appl. Phys. Lett.} \textbf{2009}, \emph{95} 144106.

\bibitem{Dyck2017}
O.~Dyck, D.~N. Leonard, L.~F. Edge, C.~A. Jackson, E.~J. Pritchett, P.~W.
  Deelman, J.~D. Poplawsky,
\newblock \emph{Adv. Mater. Interfaces} \textbf{2017}, \emph{4} 1700622.

\bibitem{Koelling2023}
S.~Koelling, L.~E.~A. Stehouwer, B.~Paquelet~Wuetz, G.~Scappucci,
  O.~Moutanabbir,
\newblock \emph{Adv. Mater. Interfaces} \textbf{2023}, \emph{10} 2201189.

\bibitem{Suedkamp2016}
T.~S{\"u}dkamp, H.~Bracht,
\newblock \emph{Phys. Rev. B} \textbf{2016}, \emph{94} 125208.

\bibitem{Kube2010}
R.~Kube, H.~Bracht, J.~L. Hansen, A.~N. Larsen, E.~E. Haller, S.~Paul,
  W.~Lerch,
\newblock \emph{J. Appl. Phys.} \textbf{2010}, \emph{107} 073520.

\bibitem{Kube2013}
R.~Kube, H.~Bracht, E.~H{\"u}ger, H.~Schmidt, J.~L. Hansen, A.~N. Larsen, J.~W.
  Ager~III, E.~E. Haller, T.~Geue, J.~Stahn,
\newblock \emph{Phys. Rev. B} \textbf{2013}, \emph{88} 085206.

\bibitem{Godbey1992}
D.~J. Godbey, M.~G. Ancona,
\newblock \emph{Appl. Phys. Lett.} \textbf{1992}, \emph{61} 2217.

\bibitem{Fukatsu1991}
S.~Fukatsu, K.~Fujita, H.~Yaguchi, Y.~Shiraki, R.~Ito,
\newblock \emph{Appl. Phys. Lett.} \textbf{1991}, \emph{59} 2103.

\bibitem{Godbey1994}
D.~J. Godbey, J.~V. Lill, J.~Deppe, K.~D. Hobart,
\newblock \emph{Appl. Phys. Lett.} \textbf{1994}, \emph{65} 711.

\bibitem{Godbey1998}
D.~Godbey, M.~Ancona,
\newblock \emph{Surf. Sci.} \textbf{1998}, \emph{395} 60.

\bibitem{Jernigan1996}
G.~G. Jernigan, P.~E. Thompson, C.~L. Silvestre,
\newblock \emph{Appl. Phys. Lett.} \textbf{1996}, \emph{69} 1894.

\bibitem{Jernigan1997}
G.~G. Jernigan, P.~E. Thompson, C.~L. Silvestre,
\newblock \emph{Surf. Sci.} \textbf{1997}, \emph{380} 417.

\bibitem{Pramann2011}
A.~Pramann, O.~Rienitz, D.~Schiel, J.~Schlote, B.~G{\"u}ttler, S.~Valkiers,
\newblock \emph{Metrologia} \textbf{2011}, \emph{48} 20.

\bibitem{Corley2023a}
C.~Corley-Wiciak, M.~Zoellner, I.~Zaitsev, K.~Anand, E.~Zatterin, Y.~Yamamoto,
  A.~Corley-Wiciak, F.~Reichmann, W.~Langheinrich, L.~Schreiber, C.~Manganelli,
  M.~Virgilio, C.~Richter, G.~Capellini,
\newblock \emph{Phys. Rev. Appl.} \textbf{2023}, \emph{20} 024056.

\bibitem{Corley2023b}
C.~Corley-Wiciak, C.~Richter, M.~H. Zoellner, I.~Zaitsev, C.~L. Manganelli,
  E.~Zatterin, T.~U. Schülli, A.~A. Corley-Wiciak, J.~Katzer, F.~Reichmann,
  W.~M. Klesse, N.~W. Hendrickx, A.~Sammak, M.~Veldhorst, G.~Scappucci,
  M.~Virgilio, G.~Capellini,
\newblock \emph{ACS Appl. Mater. Interfaces} \textbf{2023}, \emph{15} 3119.

\bibitem{Wilson1964}
D.~K. Wilson,
\newblock \emph{Phys. Rev.} \textbf{1964}, \emph{134} A265.

\bibitem{Sailer2009}
J.~Sailer, V.~Lang, G.~Abstreiter, G.~Tsuchiya, K.~M. Itoh, J.~W. Ager~III,
  E.~E. Haller, D.~Kupidura, D.~Harbusch, S.~Ludwig, D.~Bougeard,
\newblock \emph{Phys. Status Solidi RRL} \textbf{2009}, \emph{3} 61.

\bibitem{Cojocaru2013}
O.~Cojocaru-Mirédin, T.~Schwarz, P.-P. Choi, M.~Herbig, R.~Wuerz, D.~Raabe,
\newblock \emph{J. Vis. Exp.} \textbf{2013}, \emph{74} e50376.

\bibitem{Becker2010}
P.~Becker, H.-J. Pohl, H.~Riemann, N.~Abrosimov,
\newblock \emph{Phys. Status Solidi A} \textbf{2010}, \emph{207} 49.

\end{thebibliography}

\end{document}